\documentclass[prl,reprint,superscriptaddress]{revtex4-1}

\usepackage{amsmath} 
\usepackage{amssymb}  
\usepackage{amsfonts} 
\usepackage{dsfont}
\usepackage{bm}  
\usepackage{bbm}   
\usepackage{bbold}   
\usepackage{braket} 
\usepackage{color} 
\usepackage{comment}  
\usepackage{dcolumn}  
\usepackage{enumerate}  
\usepackage{epsfig}  
\usepackage{gensymb}  
\usepackage{graphicx}  
\usepackage{indentfirst}  \usepackage{lmodern}  
\usepackage{mathrsfs}  
\usepackage{mathtools}  
\usepackage{psfrag}  
\usepackage{pst-all} 
\usepackage{soul}  
\usepackage{xcolor}
\usepackage{float} 
\usepackage[colorlinks,linkcolor=blue,citecolor=blue,urlcolor=blue,hyperindex,driverfallback=dvipdfm]{hyperref}  \usepackage[T1]{fontenc} 

\usepackage{dirtytalk}
\usepackage{CJKutf8}

\usepackage{float} 
\usepackage{placeins}
\usepackage{bm}
\usepackage{enumitem}

 \usepackage[normalem]{ulem} 
 \makeatletter
\newcommand*{\addFileDependency}[1]{
  \typeout{(#1)}
  \@addtofilelist{#1}
  \IfFileExists{#1}{}{\typeout{No file #1.}}
}
\makeatother

\usepackage[strict]{changepage}
\usepackage{calc}
\usepackage{dirtytalk}
\usepackage{pifont}

\usepackage{xr-hyper}

\usepackage{graphicx}
\usepackage{cancel}
\usepackage{longtable}
\usepackage{array}
\newcolumntype{C}[1]{>{\centering\arraybackslash}p{#1}}

\DeclarePairedDelimiter\abs{\lvert}{\rvert}%

\usepackage{ulem}
\usepackage{titlesec}
 \titleformat{\paragraph}[hang]{\bfseries}{}{0pt}{\uline}
\titlespacing*{\paragraph}
{0pt}{3.25ex plus 1ex minus .2ex}{1.5ex plus .2ex}

\usepackage{dirtytalk}
\usepackage{ulem}

\begin{document}

\title{Table-Top Tunable Chiral Photonic Emitter}

\author{Lu Wang*}
\affiliation{State Key Laboratory of Magnetic Resonance and Atomic and Molecular Physics, Wuhan Institute of Physics and Mathematics, Innovation Academy for Precision Measurement Science and Technology, Chinese Academy of Sciences, Wuhan 430071, China}
\email{lu.wangTHz@outlook.com}
\author{Marcelo Fabián Ciappina}
\affiliation{Department of Physics, Guangdong Technion - Israel Institute of Technology, Shantou 515063, China}
\affiliation{Technion - Israel Institute of Technology, Haifa 32000, Israel}
\affiliation{Guangdong Provincial Key Laboratory of Materials and Technologies for Energy Conversion, Guangdong Technion - Israel Institute of Technology, Shantou 515063, China}
\author{Thomas Brabec}
\affiliation{Department of Physics, University of Ottawa, Ottawa, Ontario K1N 6N5, Canada}
\author{Xiaojun Liu}
\affiliation{State Key Laboratory of Magnetic Resonance and Atomic and Molecular Physics, Wuhan Institute of Physics and Mathematics, Innovation Academy for Precision Measurement Science and Technology, Chinese Academy of Sciences, Wuhan 430071, China}
\begin{abstract}
The increasing interest in chiral light stems from its spiral trajectory along the propagation direction, facilitating the interaction between different polarization states of light and matter. Despite tremendous achievements in chiral light-related research, the generation and control of chiral pulse have presented enduring challenges, especially at the terahertz and ultraviolet spectral ranges, due to the lack of suitable optical elements for effective pulse manipulation. Conventionally, chiral light can be obtained from intricate optical systems, by an external magnetic field, or by metamaterials, which necessitate sophisticated optical configurations. Here, we propose a versatile tunable chiral emitter, composed of only two planar Weyl semimetals slabs, addressing the challenges in both spectral ranges. Our results open the way to a compact tunable chiral emitter platform in both terahertz and ultra-violet frequency ranges. This advancement holds the potential to serve as the cornerstone for integrated chiral photonics.
\end{abstract}

\maketitle

Controlling the interplay of light and its interaction with matter has long been a fundamental objective in optics. In particular,
chiral light has attracted substantial research interests, as the light polarization spiraling along the propagation direction provides a variety of
opportunities for control and spectroscopy of matter. To name a few, chiral light is applied in subcycle optical gates such as angular-dependent dynamics studies \cite{bloch2021revealing,chen2017dynamic}, chiral molecule spectroscopy \cite{tanaka2018circularly,richardson1977circularly}, mechanical separation of chiral objects \cite{canaguier2013mechanical},  and chiral quantum optics \cite{lodahl2017chiral}.

Despite the tremendous achievements in chiral light related research, the generation and control of ultraviolet and near-infrared, chiral pulses have continued to present challenges for the following reasons. Firstly, although chiral objects are widespread in nature, such as DNA and certain proteins~\cite{mun2020electromagnetic}, the chiral light-matter interaction is typically very weak due to the large mismatch between their atomic feature sizes and optical wavelengths \cite{forbes2022chiral,mun2020electromagnetic}.  Although one can use external magnetic 
fields to enhance the material chirality, the typically required magnetic field strengths are of a few Tesla \cite{fan2021terahertz,stanciu2007all}, leading to cumbersome and costly magnetic structures that are unsuited to large-scale manufacturing.  On the other hand, metasurfaces and plasmonics \cite{hendry2012chiral,zhang2022chiral,mun2020electromagnetic} can be utilized
for manipulating chiral light. However, the necessary complex optical configurations require nano-scale fabrication with multiple 
lithographic processes \cite{santos2015low,chen2001nanofabrication}. 

Generation and control of chiral light in the ultra-violet frequency range (800\,THz-30\,PHz) is even more difficult, due to the lack 
of suitable optical elements \cite{borrego2018ultraviolet,brahms2023efficient}. This impedes progress in many areas, from high-resolution
spectroscopy \cite{perkampus2013uv} and photo-lithography \cite{del2019light}, to controlling atomic/ molecular electronic transitions 
\cite{ranitovic2014attosecond} and inactivation of microorganisms \cite{li2009assessment}.


Control and manipulation of terahertz (0.1\,THz-30\,THz) chiral pulses, on the opposite end of the optical spectrum, have proven 
equally challenging \cite{fan2021terahertz}.
Terahertz pulses can serve as an active gate to enhance the spatio-temporal resolution of spectroscopy \cite{cocker2013ultrafast}, realize relativistic electron acceleration \cite{hibberd2020acceleration}, and be the key to 6G communication \cite{rikkinen2020thz}. Nevertheless, the terahertz frequency can not be accessed easily, because it is in between the standard electronic and photonic manipulation methods  \cite{vitiello2021physics,pang2022bridging}. The difficulties in handling Terahertz pulses are mainly due to two reasons.
One is the large diffraction owing to their long wavelength nature. The second aspect involves the absorption within solids, as this frequency 
range is in resonance with lattice phonons \cite{kozina2019terahertz}. Consequently, minimizing interactions with solid-state optical 
elements is imperative to circumvent these challenges. The above difficulties have been referred to as the "terahertz gap" ~\cite{vitiello2021physics,pang2022bridging}.

\begin{figure*}[th]
\centering
\includegraphics[width=1\linewidth]{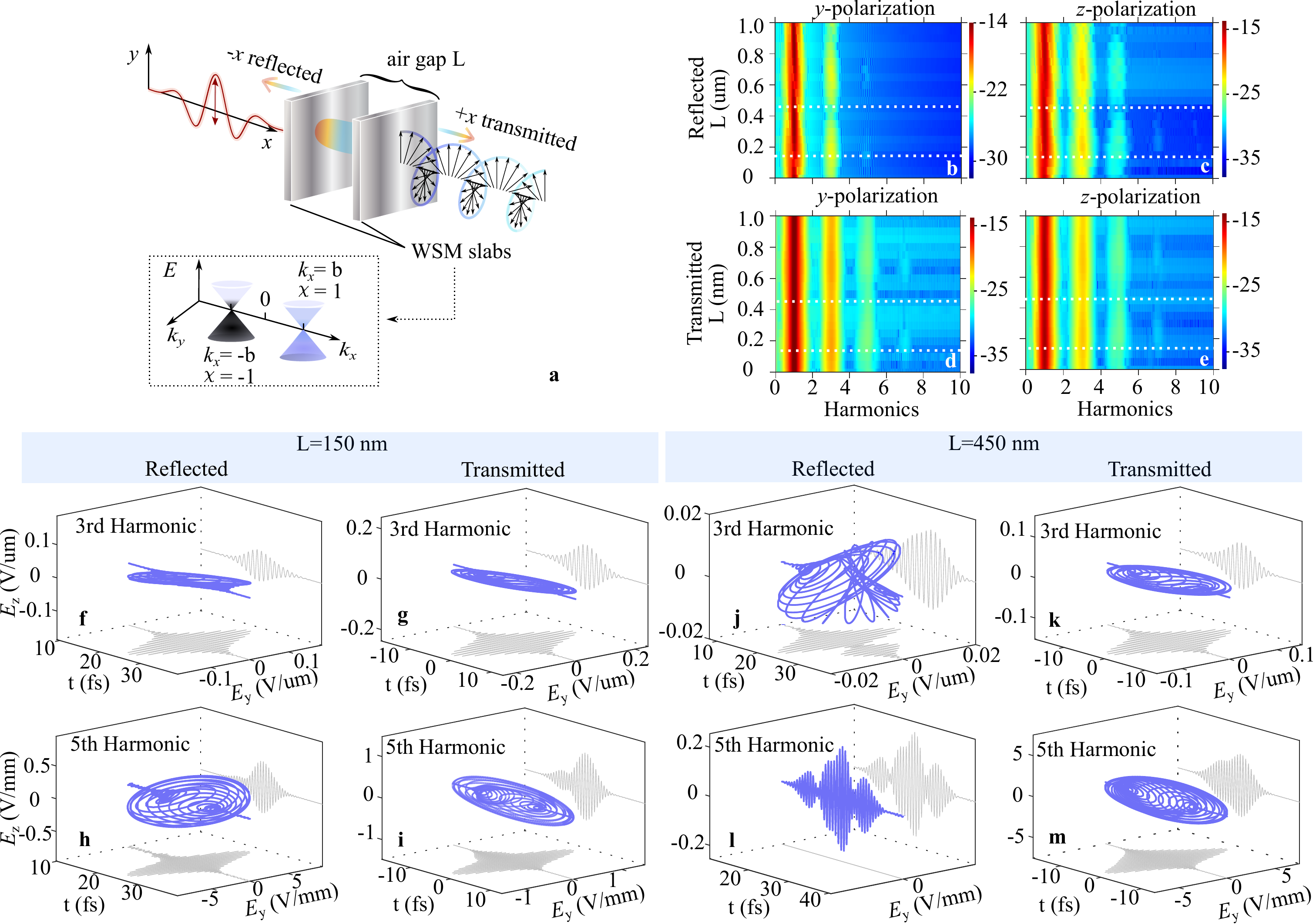}
\caption{Elliptically polarized harmonics induced by linearly polarized driving field along $y$ dimension. Panel \textbf{a} illustrates the proposed planar chiral emitter i.e. two Weyl semimetal (WSM) slabs separated by air. Panels \textbf{b,c} (\textbf{d,e}) are the reflected (transmitted) spectra of the harmonics as a function of the air separation distance $L$ for $y-$ and $z-$ polarization, respectively. The color bar represents $\log{[E(\omega)]}$. Panels \textbf{f-i} and \textbf{j-m} show the 3D temporal evolution of given harmonic orders (both transmitted and reflected) with $L=150$ nm and $L=450$ nm, respectively. These two air gap distances are indicated by horizontal dashed lines in \textbf{b-e}. The results are calculated with driving pulse parameters as, pulse duration $\tau=10$\,fs, center wavelength $\lambda_0=1\,\mu$m, and peak field strength $E_0=2\times 10^8\,\text{V}/\text{m}$ corresponding to intensity $\sim 10^{14} \text{W}/\text{m}^2$. Both the WSM slabs are of 100 nm thickness. The Weyl nodes separation is 2b along $k_x$ dimension where $b=0.06\pi/a_0$,  $a_0=3.4\r{A}$ is the lattice parameter along $x$ \cite{lv2015observation}. The Fermi energy is $100$ meV. The harmonics are at the ultra-violet spectral range.}
\label{fig:1}
\end{figure*}
In this work, we introduce a versatile tunable chiral emitter addressing the challenges in both the ultra-violet and terahertz spectral ranges at the
same time. The proposed setup employs only two planar Weyl semimetal (WSM) slabs and a linearly polarized field driving high harmonic 
generation (HHG), as illustrated in Fig.~1\textbf{a}.  Our approach utilizes the chirality of the high-harmonic generation process in
the WSM planar structures, which presents the basis for tunable chiral ultraviolet emission from a table-top source. Elimination of complex optical elements simplifies the experimental realization of the proposed setup. The theoretical model used in our analysis solves the two-band equations for semimetals and Maxwell's equations self consistently. The model applies to WSMs, Dirac semimetals, and 2D semimetals such as graphene.

WSMs are a newly discovered class of quantum materials with unique optical properties and an extremely strong nonlinearity \cite{patankar2018resonance,wu2017giant}. In the momentum space, the points where valence and conduction bands cross are Weyl nodes, around which electrons behave as massless Weyl fermions \cite{yan2017topological}. The Weyl nodes appear in pairs with opposite chirality, making them the perfect source and detector for chiral light \cite{ma2017direct,yoshikawa2022non}. An inherent feature of the chiral properties in WSMs is the intrinsic coupling of different polarization states. In other words, by choosing the right combination of the driving field's polarization direction with respect to the Weyl nodes separation direction, it is possible to achieve elliptically polarized emission using a purely linearly polarized driving field, in the absence of any external magnetic field \cite{wang2023maximal,wang2023nonreciprocal,avetissian2022high,gao2020chiral}. 

When exposed to a strong laser field, the nonlinear currents arise from the movement of electrons within the WSM through two primary pathways: the motion within the individual bands (intra-band) and the polarization accumulation between bands (inter-band) \cite{ghimire2019high,wu2015high,vampa2014theoretical}. The nonlinear current emits photons with energies that are integer multiples of the driving field. This process is known as high-harmonic generation in solids \cite{ghimire2019high,wu2015high,vampa2014theoretical}. In momentum space, the Weyl points act like magnetic monopoles. The non-vanishing Berry curvature flows from the source $\chi=1$ to the sink $\chi=-1$. These magnetic monopoles create magnetic field equivalent effects. Under the influence of the magnetic field, the electrons move toward a direction perpendicular to the applied field and velocity, resulting in harmonic emissions with more than one polarization. Details can be found in supplementary material (SM) sec. III \cite{supp}.  The generated harmonic pulses exhibit excellent coherence properties \citep{salieres1997study}. Combining the elliptical emission from the WSM and the two-slab structure, the proposed scheme realizes easy tunability of the chiral pulses via interference along the propagation direction. This is similar to the principles of a Fabry–Pérot cavity.  Moreover, since the harmonics are sensitively related to the properties of the driving field, the chiral/3D
polarization states of the high-harmonic radiation can be fully controlled through the intense pulse laser. \\

\noindent\textbf{Results.}\\
\textbf{Modifying chiral harmonic properties by tuning the distance $L$.} In Figs. 1\textbf{b-e}, the harmonic spectra, both reflected and 
transmitted, are depicted as a function of the separation distance between the two slabs, denoted as $L$.  The theoretical method 
is briefly discussed below and in more detail in the SM Sec. III \cite{supp}. The calculations are based on standard experimental available parameters. The driving field strength is $E_0=2\times 10^8 \text{V}/\text{m}$ \cite{ghimire2011observation}, the WSM slab thickness is assumed to be (100\, nm)\cite{chi2018wide,bedoya2020realization}, and standard WSM 
material parameters are used \cite{lv2015observation}. 

Since the model we chose to describe WSM breaks only the time-reversal symmetry and preserves the inversion symmetry, only odd harmonics are present (Methods sections Eqs.\ref{eq:jx}-\ref{eq:jz}). The two-slab structure forms a Fabry–Pérot interference cavity. By alternating the distances between the two WSM slabs, the boundary conditions change for the electric field (i.e. superposition of the harmonics and the driving field) interacting with the structure, resulting in modified harmonics. Visible spectral amplitude modification can be observed (note that the color bar is in log scale). The harmonic spectra are expected to exhibit periodicity concerning $L$, with a period of $\sim\lambda_0/2$, where $\lambda_0=1\, \textmu$m is the center wavelength of the driving field. The spectra observed in reflection (Figs.~1\textbf{b,c}) and transmission (Figs.~1\textbf{d,e}) manifest distinctive characteristics. For the reflected wave, the dominant emission originates directly from the immediate reflection, i.e.,~the wave reflected directly from the surface. Consequently, the spectra in reflection demonstrate weaker nonlinear effects, specifically lower-order harmonics, in contrast to the transmitted case, where more pronounced harmonic orders are evident due to the interaction of the driving pulse through two WSM slabs. It can be seen that our compact chiral light source offers the potential for generating diverse polarization states across various harmonic orders (see Fig.1\textbf{f-m}). These exotic polarization states can trigger nuanced angle-resolved electron dynamics that will enhance our understanding of light-matter interaction with a sub-cycle temporal resolution \cite{jiang2022atomic}.

\noindent\textbf{Polarization and ellipticity evolution in time.}
\begin{figure}[h]
\centering
\includegraphics[width=0.98\linewidth]{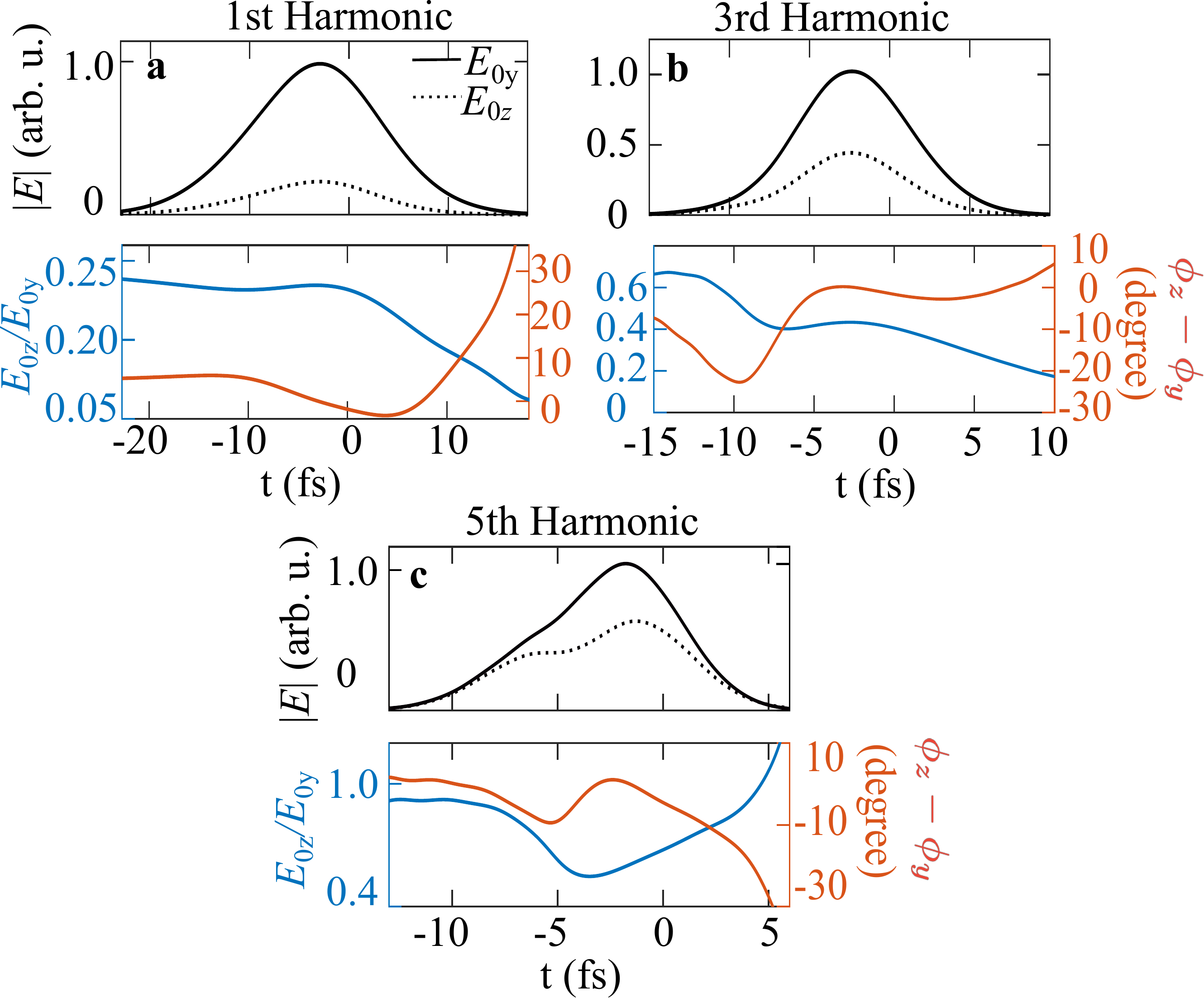}
\caption{The polarisation and ellipticity evolution for transmitted harmonics with the air gap distance $L=450\,$nm. In the upper panel, the solid and dashed curves depict the amplitude of the electric field for the specified harmonic in the $y$ and $z$ dimensions. The blue and red curves represent the ratio of the amplitude $E_{0z}/E_{0y}$, and the phase difference $\phi_z-\phi_y$, respectively. The 3D temporal evolution of \textbf{b} and \textbf{c} are presented in Figs.~1\textbf{k,m}.}
\label{fig:1}
\end{figure}
To further characterize the harmonic pulses and their temporal evolution, we examine the polarization state of a given harmonic as a function of time. When the driving pulse is polarized along the $y$ direction, the emitted harmonics are exclusively along the $y$ and $z$ directions
(SM Sec. IV \cite{supp}). The ellipticity of a pulse is determined by both the relative amplitude and phase relations of the two orthogonal polarizations. For instance, the electric field of a particular harmonic order, polarized along a given direction can be expressed as $E_i(t)=\{E_{0i}(t)\exp[i\phi_i(t)]+c.c\}/2$, where $i\in {y,z}$ and $E_{0i},\phi_i \in \Re$. For a linearly polarized pulse, it necessitates $\phi_z-\phi_y=n\pi$, where $n=0,\pm1,\pm2,...$. Similarly, a circularly polarized light demands $E_{0z}/E_{0y}=1$ and $\phi_z-\phi_y=\pi/2+n\pi$, where $n=0,\pm1,\pm2,...$. In 
Figs.~2\textbf{a-c} we depict the 1st, 3rd, and 5th harmonics, respectively, in the transmitted direction with an air gap distance $L=450\,$nm. Figures 1\textbf{k,m} and 2\textbf{b,c} represent the same electric fields. In the upper panel of Fig.~2, the solid (dashed) black curve represents the electric field envelope of $E_{0y}$ ($E_{0z}$). To quantify the pulse evolution dynamics, we define two parameters: the ratio of the amplitude $E_{0z}/E_{0y}$, and the phase difference $\phi_z-\phi_y$. These two parameters are indicated by blue and red curves on the lower panels, respectively.  The temporal evolution of the harmonic fields displays distinctive and dynamic changes throughout the entire pulse duration. In other words, for a given harmonic order, the ellipticity varies with respect to time within the pulse. {Since the multi-photon ionization is highly sensitive to the driving pulse polarisation and carrier-envelope phase, these unusual and exotic polarization changes of the generated harmonics across a single pulse can be harnessed for ultrafast electron dynamics control \mbox{\cite{eickhoff2021multichromatic,song2023control,li2021generation}}. In addition, these time-varying polarisation states of the fields may bring new symmetry to the free electron wave packets via the photon electron ionization, resulting in a powerful tool for coherent control of the quantum process \mbox{\cite{kerbstadt2019odd}}. }

\noindent\textbf{Influence of the Fermi Energy on Carrier Envelope Phase (CEP) at the Terahertz regime.} 
\begin{figure}[h]
\centering
\includegraphics[width=1\linewidth]{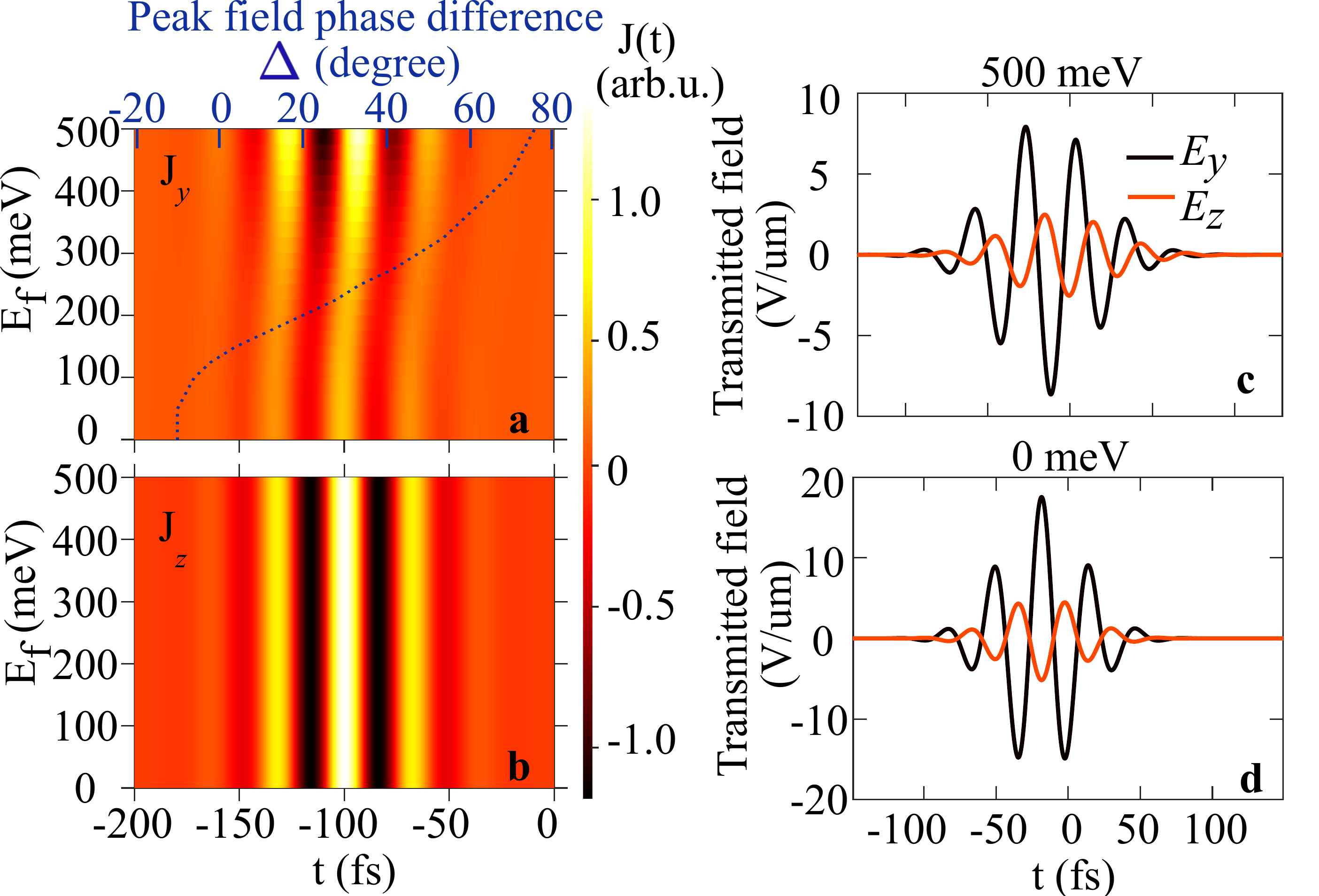}
\caption{The color plot \textbf{a, b} represent the current normalized current $J_y(t)$, $J_z(t)$, respectively. The results are calculated by $\lambda_0=10\,\textmu$m corresponds to frequency $f=10\,$THz and $\hbar2\pi f\approx 120\,$meV, pulse duration $\tau=40\,$ fs, electric field is polarized along $\hat{y}$ with peak field strength $E_0=2\times10^7 V/m$. The transmitted electric fields after two Weyl semimetal slabs (100\,nm each) separated by 2\textmu m in the air are presented in \text{c,d} where the Fermi energies are $0$ and $500$ meV, respectively.}
\label{fig:1}
\end{figure}
The proposed scheme in Fig.~1\textbf{a} is also valid for the terahertz frequency range. Thus, we will not repeat the same calculations. Instead, we analyze how the waveform of the terahertz pulse varies as a function of the Fermi energy $\text{E}_\text{f}$.
Since terahertz pulses are frequently applied to electron acceleration and communication, the capacity to tailor their waveform is very important.
With increasing Fermi energy, the inter-band transition is suppressed, so that the intra-band transition starts to dominate. Since the inter- and intra-band transitions are not always in phase, the final induced current has a phase shift (see Figs.~3\textbf{a,b}). The peak phase differences 
of the current are defined as $\Delta=(t_{y}-t_z)\omega_0$ where $t_i$ is the time corresponding to the peak of the electric field polarized along the $i$ direction with $i\in\{y,z\}$. The value of $\Delta$ as a function of $\text{E}_\text{f}$ is plotted  as the dotted curve in 
Fig.~3\textbf{a} with the horizontal axis on top. The transmitted electric fields with $\text{E}_\text{f}=0$ and $\text{E}_\text{f}=500$ meV are presented in Fig.~3\textbf{c} and \textbf{d}, respectively. The two Weyl semimetal slabs (100\,nm each) are separated by 2\textmu m in the air.

\noindent\textbf{Discussion and Conclusion}\\
Several WSMs have been experimentally discovered, such as  $\text{Y}_2\text{Ir}_2\text{O}_7$,  $\text{HgCr}_2\text{Se}_4$, $\text{TaAs}$, TaP and $\text{Co}_3\text{Sn}_2\text{S}_2$ \mbox{\cite{kumar2017extremely,wan2011topological,xu2011chern,yan2017topological,okamura2020giant}}, which could be used for an experimental implementation of our scheme. Furthermore, previous experiments have successfully grown WSMs from nanostructures \cite{bachmann2017inducing} to crystals, grown by the chemical vapor transfer method \cite{ma2017direct}. They are accessible within a considerable size range, spanning a few nanometers \cite{bedoya2020realization} to millimeters \cite{chi2018wide}. 

The compact chiral light source we propose opens the door to various applications. For example, it creates a new dimension for information coding in optical communication \mbox{\cite{forbes2022chiral}}. Besides, since chiral molecules respond differently to chiral light, chiral light sources can be used to sort molecules and achieve all-optical chiral discrimination \mbox{\cite{genet2022chiral,zhang2017all}}. Furthermore, this easily tunable setup can be integrated with chemical experiments and serve as a sensitive trigger for photochemical reactions \mbox{\cite{rikken2000enantioselective}}. 

{Our methodology also has potential applications in the realm of band structure spectroscopy of WSMs. For example, the chirality of the emitted harmonics is determined by the Weyl nodes separation $b$, given the fact that the emission is only linearly polarized if $b=0$ (see Method section and SM Sec. IV \mbox{\cite{supp}}). As a result, emissions observed along the $\hat{z}$ direction become a pivotal metric for deducing the Weyl nodes' separation, $b$. Besides, since the nonlinearity of the emitted harmonics is proportional to the Fermi velocities \mbox{\cite{wang2022highly}} as shown in Eqs. (2-4), the Fermi velocities of WSMs can be retrieved by obtaining the amplitude of a given harmonic. }

\noindent\textbf{Method.} To achieve the proposed objective mentioned above, we choose the most straightforward model to describe the WSM. In this approach, the time-reversal symmetry is broken whereas the inversion symmetry is preserved. We focus on a two-level system with one pair of Weyl nodes of opposite chirality separated by a distance 2$b$ along the $\hat{k}_{x}$ dimension in the momentum space.  Specifically, the Hamiltonian is written as \cite{avetissian2022high,hasan2017discovery,armitage2018weyl,ishikawa2010nonlinear,lim2020efficient}:
\begin{equation}
    H_\chi=\chi\sum_{i\in\{x,y,z\}}v_i\sigma_i\pi_i,\quad \chi=\pm 1,
\end{equation}
where $x,\, y,\,z$ are Cartesian coordinates, $v_x=v_y=v_z=1\times 10^6\,\text{m}/\text{s}$ are the Fermi velocities \cite{armitage2018weyl}, $\sigma_i$ with $i\in\{x,\,y,\,z\}$ define the Pauli matrices, and $\chi$ represents the chirality of the node. The kinetic momentum of the electron under the influence of an external field (for a detailed derivation see SM Sec.~III \cite{supp}) is described by $\pi_i=\hbar(k_i-\chi b_i)-qA_i$, where $\hbar k_i$ is the electron momentum in the absence of an external field, $q=-e$, with $e=\abs{e}$ the elementary charge, and $\bm{b}=(b,0,0)$ is the node off-set with respect to the origin in the momentum space. The vector potential is defined as $A_i=-\int_{-\infty}^tE_i(t^\prime)dt^\prime$, where $E_i$ is the electric field polarized along $i$. The total current along the $i$ direction takes into account the contributions from both nodes, i.e.~$J_i(t)=J_{i}(\chi=+1,t)+J_{i}(\chi=-1,t), i\in\{x,y,z\}$ \cite{avetissian2022high}.

{The general form of the current can be found in SM Sec. III Eqs. S34-S36 \mbox{\cite{supp}}. These expressions are highly comprehensive. Setting $b=0$ results in the expression for Dirac semimetals \mbox{\cite{lim2020efficient}}. Based on this, by further eliminating the third dimension $z$, one obtains the expression for graphene \mbox{\cite{ishikawa2010nonlinear}}.

In our configuration, the Weyl nodes separation is along $\hat{x}$ and the driving field is linearly polarized along $\hat{y}$. Applying the frozen band approximation and accounting for k-space symmetry (refer to SM Sec. IV \cite{supp}), we simplify the analytical expressions for the current as the following:}
\begin{align}
    &j_x(t)=0\label{eq:jx}\\
    &j_y(t)=-2\chi\mathcal{M}_0\frac{\Omega_0}{\omega_0}\text{Im}\left[\sum_{\substack{n,m=-\infty\\
    n-m\in \text{ odd}}}^\infty\hspace{-4mm} \mathcal{C}_2(N,n){C}_3(N,m)^*\large\right]\label{eq:jy}\\
    &j_z(t)=\frac{\sin(\theta)}{\cos(
\phi)}\mathcal{M}_0\sum_{\substack{n,m=-\infty\\
    n-m\in \text{ odd}}}^\infty\mathcal{C}_1(N,n)\mathcal{C}_1(N,m)^*\nonumber\\
    &-2vq\frac{\sin(\theta)^2}{\cos(
    \theta)}  \text{Re}\left[\sum_{\substack{n,m=-\infty\\
    n-m\in \text{ odd}}}^\infty\mathcal{C}_1(N,n)\mathcal{C}_3(N,m)^*\right].\label{eq:jz}
\end{align}
The related variables are defined as
\begin{align}
   & \mathcal{C}_1(N,n)=\frac{n\text{J}_n(N)\exp(i\omega_0n)}{n+\Omega_0/\omega_0}\exp(-\frac{t^2}{\tau^2}),\nonumber\\
   & \mathcal{C}_2(N,n)=\frac{\text{J}_n(N)\exp(i\omega_0n)}{n+\Omega_0/\omega_0}\exp(-\frac{t^2}{\tau^2}),\nonumber\\
   &\mathcal{C}_3(N,n)={\text{J}_n(N)\exp(i\omega_0n)},\nonumber\\
& \mathcal{M}_0=qv\frac{\sin(\theta)[\cos(\phi)^2+\sin(\phi)^2\cos(\theta)^2]}{\cos(\phi)\cos(\theta)},\nonumber\\
&N=\frac{-\chi q\exp(-{t}^2/\tau^2)\cos{\phi}\cos{\theta}E_{0}}{({2\omega_0\hbar|k|})}\nonumber
\end{align}
{where $\text{J}_n$ is the Bessel function of the first kind, $\Omega_0=4v|q|E_{0}/\hbar\pi\omega_0$, $\theta=\arccos (k_z/|k|)$, $|k|=\sqrt{(k_x-\chi b)^2+k_y^2+k_z^2}$, and $\phi=\arctan[k_y / (k_x-\chi b)]$. It can be seen that only odd harmonics survive (see SM Sec. IV for details). In particular, Eq.(\mbox{\ref{eq:jz}}) is proportional to the Weyl nodes separation $b$. If the two Weyl nodes of the opposite chirality merge to one place in k space (i.e. $b=0$), $j_z(t)=0$ [see SM Sec. IV Eq.(S84)]. }

The planar structure of the WSM slabs is modeled via a customized Finite-Difference Time-Domain (FDTD) method. The emission current (we have used the general form as shown in Eqs. S34-S36) is integrated into the Maxwell equations. The numerical details are presented in SM Secs. VI and VII \cite{supp}. The total current requires integration over the 3D $k$-space. The most standard mesh point selection would be an even-distance mesh, where up to 500 mesh points along each dimension are typically required \cite{avetissian2022high}. We choose the parameters for one of the most standard Weyl semimetals i.e.~TaAs. The lattice parameters at 300 K are $a_0=b_0=3.4\r{A}$ and $c_0=11.6\r{A}$ \cite{lv2015observation}.  We choose to integrate over the entire Brillouin space with the integration range as $[-\pi/r,\pi/r]$, where $r\in\{a_0,b_0,c_0\}$. For very large $k$ values, there is almost no transition \cite{ishikawa2010nonlinear}. This can be understood intuitively. With a very large $k_z$ for example, in the $k_x-k_y$ plane, the Weyl semimetal behaves as a material with a very large band gap ($\sim 2\hbar v_z \abs{k_z}$). Thus, the inter-band transitions are harder. Since the dominant inter-band transitions are around the Weyl nodes, in the entire Brillouin zone, many mesh points can be ignored, i.e., the current is almost zero in regions far from the Weyl nodes (SM Sec.~V \cite{supp}). Consequently, we use non-equal-distance mesh sizes, which are denser around the cone (SM Sec.~V \cite{supp} Fig.S7). With this approach, we obtained comparable results as  \cite{avetissian2022high} (details see Sec.~VI  Fig. S9) using a mesh size of $60\times 60\times 60$, which is $\sim 10^3$ times smaller than the conventional method \cite{avetissian2022high,vampa2014theoretical}. For our work, one single job of the high-harmonic generation via the 2 Weyl semimetal slabs (mesh size 60$\times60\times60$) requires a computational time of $\sim12$ hours on a single node with 128-core CPU @ 2.6GHz.\\
%

\noindent\textbf{Data availability}\\
Data supporting the findings of this work are available from the corresponding authors upon reasonable request.
\noindent\textbf{Code availability}\\
The code developed for this work is available to the public at the repository:\\
\noindent\textbf{Acknowledgements}\\ We would like to thank Beijing Beilong Super Cloud Computing Co., Ltd for their support in HPC. In particular, L.W. thanks Mr. X.L. Han, Mr. R. Liu, and Mr. H.B. Chen for their timeless help. This work is supported by the National Key Research and Development Program of China (Grant No. 2019YFA0307702) and the National Natural Science Foundation of China (Grant No. 12121004). M.F.C acknowledges financial support from the Guangdong Province Science and Technology Major Project (Future functional materials under extreme conditions - 2021B0301030005) and the Guangdong Natural Science Foundation (General Program project No. 2023A1515010871). \\
\noindent\textbf{Author contributions}\\ 
\noindent\textbf{Competing interests}\\ 
The authors declare no competing interests.\\
\noindent\textbf{Additional information}\\
Supplementary information: The online version contains supplementary
material available at \\
\textbf{Correspondence and requests for materials} should be addressed to Lu Wang.
\bibliographystyle{apsrev4-1} 
\bibliography{apssamp} %
\end{document}